\documentclass[preprint,preprintnumbers]{revtex4}

\usepackage{graphicx}
\usepackage{dcolumn}
\usepackage{bm}
\usepackage{epsfig}
\usepackage{epstopdf}
\usepackage{grffile}
\usepackage{color}
\usepackage{colordvi}
\usepackage{array}
\usepackage{tikz} 
\usepackage{tikz-feynman}
\usepackage{amsmath}
\usepackage{amssymb}
\usepackage{rotating}
\usepackage{lscape}
\usepackage{float}
\usepackage{footnote}
\usepackage{wasysym}
\usepackage{gensymb}
\usepackage{subcaption}
 \makesavenoteenv{environmentname}

\usepackage{hyperref}
\usepackage{ragged2e}
\makeatletter
\long\def\@makecaption#1#2{%
	\vskip\abovecaptionskip
	\sbox\@tempboxa{#1: #2}%
	\ifdim \wd\@tempboxa >\hsize
	#1: \justifying #2\par
	\else
	\global \@minipagefalse
	\hb@xt@\hsize{\hfil\box\@tempboxa\hfil}%
	\fi
	\vskip\belowcaptionskip}
\makeatother

\begin{document}


\title{Asymmetric Two-Component Scalar FIMP Dark Matter}

\author{S. Peyman Zakeri}
\email{peymanzakeri@gmail.com}
\affiliation{Faculty of Physics, Yazd University, P.~O.~Box 89195--741, Yazd, Iran}

\date{\today}

\begin{abstract}
We propose a two-component asymmetric feebly interacting massive particle (FIMP) dark matter (DM) model in which both DM candidates are complex scalar fields. The model is an extension of the standard model (SM) by two scalar DM components and a heavy scalar mediator, stabilized by a $\mathbb{Z}_2 \times \mathbb{Z}_2'$ symmetry. DM is produced via the freeze-in mechanism through the Higgs portal, while the CP-violating decay of the heavy mediator generates an asymmetry in the first component, which is partially transferred to the second component via quartic interactions. By solving the Boltzmann equations numerically, we compute the relic density and perform a detailed scan over the parameter space. The observed relic density $\Omega_{\text{DM}} h^2 = 0.12 \pm 0.001$ is successfully reproduced for benchmark parameters $m_{\phi_1}=0.1$~GeV, $m_{\phi_2}=0.5$~GeV, $M_S=100$~GeV, and $\mu_{12}=3\times10^{-9}$, with the second component contributing only about $5\%$ to the total abundance. We also examine phenomenological constraints. The DM self-interaction cross section lies orders of magnitude below the Bullet cluster bound ($\sigma/m < 0.47$~cm$^2$/g) and the more stringent double radio relic limit ($\sigma/m < 0.22$~cm$^2$/g). The invisible Higgs decay branching ratio is $\sim 6\times10^{-18}$, well below the LHC upper limit, and direct detection prospects are negligible due to the small Higgs portal couplings. Our model establishes a novel connection between two-component DM, the freeze-in mechanism, and DM asymmetry.
\end{abstract}

\maketitle


\section{Introduction}\label{sec:int} 
The existence of DM is now firmly established by multiple independent observations. From flat galactic rotation curves~\cite{Rubin:1970} to the mass distribution in the Bullet Cluster~\cite{Clowe:2006} and the precise measurements of the cosmic microwave background (CMB) by Planck~\cite{Planck:2020}, the gravitational evidence for DM is overwhelming. These observations indicate that DM constitutes about 85\% of the total matter content of the Universe. Despite this, the particle nature of DM remains completely unknown.

The most studied hypothesis for DM is the weakly interacting massive particle (WIMP), which naturally reproduces the observed relic abundance through the thermal freeze-out mechanism~\cite{Zeldovich:1965,Chiu:1966}. In this paradigm, DM particles are initially in thermal equilibrium with the SM plasma and decouple when the expansion rate exceeds the annihilation rate. However, after decades of intensive searches, including direct detection experiments such as LUX-ZEPLIN (LZ) and XENONnT, as well as collider searches at the LHC, no conclusive WIMP signal has been found~\cite{LZ:2023,XENON:2023}. This null result, combined with the absence of any clear evidence for new physics at the TeV scale, motivates exploring alternative production mechanisms for DM.

An attractive alternative to the WIMP paradigm is the freeze-in mechanism, where DM particles never reach thermal equilibrium with the SM bath due to their extremely feeble couplings~\cite{Hall:2010,Bernal:2017}. In this scenario, often referred to as FIMPs, the DM abundance is produced gradually through the decays or scatterings of bath particles. The final relic density is determined by the small coupling constants and is typically proportional to their square~\cite{Ayazi:2015}. For a detailed discussion of the freeze-in mechanism and its applications, see also recent reviews~\cite{Belanger:2023}. The freeze-in framework has garnered significant attention as it naturally accommodates very weakly coupled DM candidates that evade direct detection limits.

An intriguing possibility that has gained significant attention in recent years is that DM may not be a single particle but rather consists of two (or more) distinct components~\cite{Profumo:2009,Bhattacharya:2013,Esch:2014,CarvalhoCorrea:2025,Zakeri:2018}. Such multi-component scenarios arise naturally in various extensions of the SM, including those with $\mathbb{Z}_N$ symmetries or hidden sector dynamics~\cite{Z5model:2020,DoubletDM:2020}. Two-component DM frameworks offer richer phenomenology compared to their single-component counterparts, allowing for novel production mechanisms, distinct detection signatures, and the possibility of addressing multiple observational constraints simultaneously~\cite{Chatterjee:2025,Kumar:2024}. Both freeze-out and freeze-in realizations of two-component DM have been extensively studied in the literature, as have models where the second component helps to resolve small-scale structure anomalies~\cite{Tulin:2018,Yang:2025} or contributes to the cosmic baryon asymmetry~\cite{Bhattacharya:2025}.

Another long-standing puzzle in cosmology is the striking similarity between the DM and baryon abundances. Observations indicate that $\Omega_{\text{DM}}/\Omega_B \approx 5$, a coincidence that any complete model of DM should ideally explain~\cite{Nussinov:1985,Petraki:2013,Becker:2019}. In the standard WIMP paradigm, the DM relic density is determined solely by its annihilation cross section, independent of the baryon asymmetry. An appealing alternative is asymmetric DM (ADM), where the DM particle carries a conserved charge analogous to baryon number~\cite{Kapustin:1998,Dutta:2011}. In ADM scenarios, the DM abundance is set by an asymmetry between particles and antiparticles, much like the visible baryon asymmetry. Consequently, the DM mass is naturally predicted to be $m_{\text{DM}} \approx 5$~GeV in minimal realizations, though extensions can accommodate heavier or multiple components~\cite{Ritter:2024,Bodas:2024}. Models where the DM asymmetry is generated simultaneously with the baryon asymmetry, often referred to as co-genesis scenarios, provide a dynamical explanation for the observed coincidence~\cite{Becker:2019,Falkowski:2011,Shigekami:2026}.

In this paper, we propose a novel extension of the FIMP framework by combining three key features: two-component DM, scalar DM candidates, and an asymmetry. Specifically, we propose a model featuring two complex scalar DM fields, whose stability is ensured by a $\mathbb{Z}_2 \times \mathbb{Z}_2'$ symmetry, and whose abundance is generated via freeze-in through the Higgs portal. A heavy scalar mediator is added to the dark sector, whose out-of-equilibrium decay generates an asymmetry in the first DM component \cite{Unwin:2014,Goudelis:2022,Asai:2023,Zakeri:2026}. This asymmetry is then partially transferred to the second component via their quartic interaction. Unlike conventional two-component DM models without an asymmetry, the present model offers a richer phenomenology. We find that the model successfully reproduces the observed DM abundance while satisfying other experimental constraints. In particular, we show that the self-interaction cross section remains safely below the Bullet cluster limit for the entire viable parameter space, and the model evades direct detection and the invisible Higgs decay constraints.

This paper is organized as follows. In Section~\ref{sec:model}, we present the model setup, including the particle content, symmetries, and the Lagrangian. Section~\ref{sec:boltz} is devoted to the asymmetric freeze-in production mechanism and the Boltzmann equations. In Section~\ref{sec:result}, we present our numerical results for the relic density and the parameter space scan. Section~\ref{sec:pheno} discusses the phenomenological constraints, namely DM self-interactions and invisible Higgs decays. Finally, we conclude in Section~\ref{sec:conclud}.

\section{Model Setup}\label{sec:model}
\subsection{Particle Content and Symmetries}
We extend the SM by introducing three scalar fields: two complex scalar $\phi_1$ and $\phi_2$, and one real scalar $S$. All new fields are singlets under the SM gauge group $SU(3)_c \times SU(2)_L \times U(1)_Y$, and therefore carry no SM quantum numbers. This guarantees that they do not receive any tree-level couplings to the SM gauge bosons. To ensure the stability of both DM components, we impose a discrete $\mathbb{Z}_2 \times \mathbb{Z}_2'$ symmetry. The transformation properties of the fields under this symmetry are summarized in Table~\ref{tab:symmetries}.

\begin{table}[h]
	\centering
	\begin{tabular}{|c|c|c|}
		\hline
		Field & $\mathbb{Z}_2$ & $\mathbb{Z}_2'$ \\
		\hline
		$\phi_1$ & $-$ & $+$ \\
		$\phi_2$ & $+$ & $-$ \\
		$S$ & $+$ & $+$ \\
		$H$ (SM Higgs) & $+$ & $+$ \\
		SM fermions & $+$ & $+$ \\
		\hline
	\end{tabular}
	\caption{Charge assignments of the new fields under the $\mathbb{Z}_2 \times \mathbb{Z}_2'$ symmetry. A $-$ sign indicates odd parity, while $+$ indicates even parity.}
	\label{tab:symmetries}
\end{table}

Under this symmetry assignment, $\phi_1$ is odd only under $\mathbb{Z}_2$, $\phi_2$ is odd only under $\mathbb{Z}_2'$, while $S$ and all SM fields are even under both. Consequently, the lightest odd particle charged under each $\mathbb{Z}_2$ is absolutely stable. Since $\phi_1$ and $\phi_2$ are the only fields odd under their respective symmetries, they cannot decay into SM particles or into each other. This simultaneously guarantees the stability of both DM candidates. The mediator $S$, being even under both symmetries, can couple to both the SM Higgs and the DM fields, thus providing a bridge between the visible and dark sectors.

In addition to the discrete symmetries, we allow for an explicit breaking of a global $U(1)$ dark number through a soft trilinear term $\mu_{12} S \phi_1^2 + \text{h.c.}$ in the Lagrangian. This term generates the asymmetry through CP-violating decays of the mediator $S$. The field content therefore consists of two complex scalar fields ($\phi_1$, $\phi_2$) as the DM candidates and one real scalar field ($S$) as the heavy mediator.

\subsection{Lagrangian}

The total Lagrangian of the model is written as
\begin{equation}
	\mathcal{L} = \mathcal{L}_{\text{SM}} + \mathcal{L}_{\text{DM}} + \mathcal{L}_{\text{Mediator}} + \mathcal{L}_{\text{Portal}} + \mathcal{L}_{\text{Asymmetry}},
\end{equation}
where $\mathcal{L}_{\text{SM}}$ is the SM Lagrangian. The DM sector consists of two complex scalar fields $\phi_1$ and $\phi_2$ with the following kinetic and mass terms
\begin{equation}
	\mathcal{L}_{\text{DM}} = |\partial_\mu \phi_1|^2 + |\partial_\mu \phi_2|^2 - m_1^2 |\phi_1|^2 - m_2^2 |\phi_2|^2 - \lambda_1 |\phi_1|^4 - \lambda_2 |\phi_2|^4 - \lambda_{12} |\phi_1|^2 |\phi_2|^2.
\end{equation}
The heavy mediator $S$ is a real scalar with its own kinetic and potential terms
\begin{equation}
	\mathcal{L}_{\text{Mediator}} = |\partial_\mu S|^2 - M_S^2 |S|^2 - \lambda_S |S|^4.
\end{equation}
The portal interactions connecting the dark sector to the SM are given by
\begin{equation}
	\mathcal{L}_{\text{Portal}} = -\lambda_{1H} |\phi_1|^2 |H|^2 - \lambda_{2H} |\phi_2|^2 |H|^2 - \lambda_{SH} |S|^2 |H|^2 - \lambda_{1S} |\phi_1|^2 |S|^2 - \lambda_{2S} |\phi_2|^2 |S|^2.
\end{equation}
The explicit breaking of the dark $U(1)$ number, which generates the asymmetry, is achieved through the trilinear term
\begin{equation}
	\mathcal{L}_{\text{Asymmetry}} = -\left( \mu_{12} S \phi_1^2 + \mu_{12}^* S^* \phi_1^{*2} \right),
\end{equation}
where the complex phase of $\mu_{12}$ leads to CP violation. In our numerical analysis, we treat the magnitude $|\mu_{12}|$ as a free parameter and effectively account for the CP-violating phase by introducing a difference in the decay widths $\Gamma_{S\to \phi_1\phi_1}$ and $\Gamma_{S\to \bar{\phi}_1\bar{\phi}_1}$.

All couplings in the above expressions are assumed to be real, except for $\mu_{12}$ which is taken to be complex to accommodate CP violation. After electroweak symmetry breaking, the SM Higgs doublet acquires a vacuum expectation value $v = 246$ GeV, and we write $H = (0, (v+h)/\sqrt{2})^T$, where $h$ is the physical Higgs boson with mass $m_h = 125$ GeV. The portal couplings then generate mixing between the scalar fields. However, since we focus on the freeze-in regime where all couplings to the SM are extremely small, we neglect such mixings in the following analysis. For completeness, we note that the bare masses $m_1$ and $m_2$ receive contributions from the $\lambda_{iH}$ couplings: $m_{\phi_i}^2 = m_i^2 + \lambda_{iH} v^2 / 2$. In the numerical analysis, we treat $m_{\phi_i}$ as the physical masses.

\section{Asymmetric Freeze-in Production}\label{sec:boltz} 

\subsection{Boltzmann Equations}\label{sec3.1}
To compute the DM relic abundance, we need to solve the Boltzmann equations that describe the evolution of the number densities of the DM candidates. For a particle species $i$, the number density $n_i$ is related to the yield $Y_i = n_i / s$, where $s$ is the entropy density of the Universe. It is convenient to work with the dimensionless variable $x = m_{\phi_1} / T$, where $m_{\phi_1}$ is the mass of the lighter DM component and $T$ is the temperature. The Hubble parameter during radiation domination is
\begin{equation}
	H(x) = \sqrt{\frac{\pi^2 g_*}{90}} \frac{m_{\phi_1}^2}{M_{\text{Pl}}} \frac{1}{x^2},
\end{equation}
where $g_* \approx 106.75$ is the effective number of relativistic degrees of freedom and $M_{\text{Pl}} = 1.22 \times 10^{19}$~GeV is the Planck mass. The entropy density is given by
\begin{equation}
	s(x) = \frac{2\pi^2}{45} g_{*s} \left( \frac{m_{\phi_1}}{x} \right)^3,
\end{equation}
with $g_{*s} \approx 106.75$ the effective degrees of freedom for entropy.

In the freeze-in regime, the DM particles are never in thermal equilibrium with the SM bath due to their extremely feeble couplings. This implies that their yields are always much smaller than the equilibrium values, $Y_i \ll Y_i^{\text{eq}}$, and the inverse annihilation processes can be safely neglected. The dominant production mechanism for both $\phi_1$ and $\phi_2$ is the decay of the SM Higgs boson $h$ into pairs of DM particles. The decay width for $h \to \phi_i \phi_i$ is
\begin{equation}
	\Gamma_{h \to \phi_i \phi_i} = \frac{\lambda_{iH}^2 v^2}{32\pi m_h} \sqrt{1 - \frac{4m_{\phi_i}^2}{m_h^2}}.
\end{equation}
Since $m_{\phi_i} \ll m_h$ in our parameter space, the phase space factor is approximately unity. 

The production of $\phi_i$ from Higgs decays is described by the following contribution to the Boltzmann equation
\begin{equation}
	\frac{dY_{\phi_i}}{dx} = \frac{1}{x H(x)} \frac{\Gamma_{h \to \phi_i \phi_i}}{s(x)} \frac{K_1(x\, m_h/m_{\phi_i})}{K_2(x\, m_h/m_{\phi_i})} Y_h^{\text{eq}}(x),
\end{equation}
and similarly for the antiparticle $\bar{\phi}_i$. For the first component, however, the production from the mediator decay generates an asymmetry, as discussed below. Here, $K_1$ and $K_2$ are modified Bessel functions of the second kind, and $Y_h^{\text{eq}}(x)$ is the equilibrium yield of the Higgs boson
\begin{equation}
	Y_h^{\text{eq}}(x) = \frac{45}{2\pi^4} \frac{g_h}{g_{*s}} \left( \frac{m_h}{m_{\phi_1}} \right)^2 x^2 K_2\!\left( x \frac{m_h}{m_{\phi_1}} \right),
\end{equation}
with $g_h = 1$ the number of degrees of freedom of the Higgs boson. For $\phi_2$, the same Boltzmann equation applies with $\lambda_{1H} \to \lambda_{2H}$, while the Bessel functions and $Y_h^{\text{eq}}(x)$ remain defined with $m_{\phi_1}$ since $x = m_{\phi_1}/T$ is kept fixed.

For the asymmetric component, we need to track the evolution of the particle and antiparticle separately. The asymmetry in $\phi_1$ is defined as $\eta(x) = Y_{\phi_1}(x) - Y_{\bar{\phi}_1}(x)$. In the freeze-in regime, the production rates for $\phi_1$ and $\bar{\phi}_1$ from Higgs decays are identical. However, the trilinear term $\mu_{12} S \phi_1^2 + \text{h.c.}$ in the Lagrangian leads to a CP-violating decay of the mediator $S$, which generates a non-zero $\eta$ during the evolution. The initial condition at high temperature is
\begin{equation}\label{eq.11}
	Y_{\phi_1}(x_{\text{in}}) = 0, \qquad Y_{\bar{\phi}_1}(x_{\text{in}}) = 0, \qquad Y_{\phi_2}(x_{\text{in}}) = 0, \qquad Y_{\bar{\phi}_2}(x_{\text{in}}) = 0,
\end{equation}
and the asymmetry is computed from the solution of the Boltzmann equations. The total yield for each component is then obtained by integrating the Boltzmann equations from $x_{\text{in}} = m_{\phi_1} / 1000$ to $x_{\text{out}} = 100$, which corresponds to temperatures well below the DM mass.

\subsection{Relic Density}\label{sec3.2}
After solving the Boltzmann equations, we obtain the final yields $Y_{\phi_i}^{\text{final}} = Y_{\phi_i}(x_{\text{out}})$ and $Y_{\bar{\phi}_i}^{\text{final}} = Y_{\bar{\phi}_i}(x_{\text{out}})$ for each component. The total yield for each DM particle is the sum of particle and antiparticle contributions, $Y_{\phi_i}^{\text{total}} = Y_{\phi_i}^{\text{final}} + Y_{\bar{\phi}_i}^{\text{final}}$. The relic density of the Universe contributed by each component is then given by
\begin{equation}
	\Omega_i h^2 = 2.755 \times 10^8 \left( \frac{m_{\phi_i}}{\text{GeV}} \right) Y_{\phi_i}^{\text{total}},
\end{equation}
where $h$ is the reduced Hubble parameter in units of $100\ \text{km}\,\text{s}^{-1}\,\text{Mpc}^{-1}$. The total DM relic density is therefore the sum of the two components
\begin{equation}
	\Omega_{\text{DM}} h^2 = \Omega_1 h^2 + \Omega_2 h^2.
\end{equation}
This quantity has been measured with high precision by the Planck collaboration, yielding the value \cite{Planck:2020}
\begin{equation}
	\Omega_{\text{DM}} h^2 = 0.120 \pm 0.001.
\end{equation}

In our analysis, we treat $m_{\phi_1}$, $m_{\phi_2}$, $M_S$, $\lambda_{1H}$, $\lambda_{2H}$, and the CP-violating parameter $\mu_{12}$ as free parameters. The Higgs portal couplings $\lambda_{1H}$ and $\lambda_{2H}$ are taken to be small enough to ensure that the freeze-in mechanism is at work. The masses are chosen in the sub-GeV range to allow for the Higgs decay channel to be kinematically accessible, while the mediator mass $M_S$ is taken to be sufficiently heavy ($M_S \gg m_{\phi_i}$) to suppress its direct production. The asymmetry is dynamically generated through the CP-violating decays of the mediator $S$, and is computed from the solution of the Boltzmann equations. The final asymmetry $\eta(x_{\text{out}}) = Y_{\phi_1}(x_{\text{out}}) - Y_{\bar{\phi}_1}(x_{\text{out}})$ is then obtained as a result of the numerical integration.

For a fixed set of parameters, the relic density is primarily determined by the CP-violating parameter $\mu_{12}$, the mediator mass $M_S$, and the mass of the first component $m_{\phi_1}$. In the next section, we present the numerical results of our scans over the parameter space, showing the temperature evolution of the yields and the dependence of the relic density on these parameters.

\section{Numerical Results and Discussion}\label{sec:result}
\subsection{Abundance Evolution}\label{sec4.1}
We now present the numerical results obtained by solving the Boltzmann equations described in Sec.~\ref{sec:boltz}. Throughout this section, we fix the masses to $m_{\phi_1}=0.1$~GeV and $m_{\phi_2}=0.5$~GeV, the Higgs portal couplings to $\lambda_{1H}=7\times10^{-11}$ and $\lambda_{2H}=7\times10^{-12}$, and the mediator mass to $M_S=100$~GeV. The CP-violating parameter $\mu_{12}$ is chosen as $3\times10^{-9}$, which yields a final asymmetry $\eta(x_{\text{out}}) \approx 0.05$ from the numerical solution.

\subsubsection{Dependence on $\mu_{12}$}

\begin{figure}[htbp]
	\centering
	\includegraphics[width=0.7\textwidth]{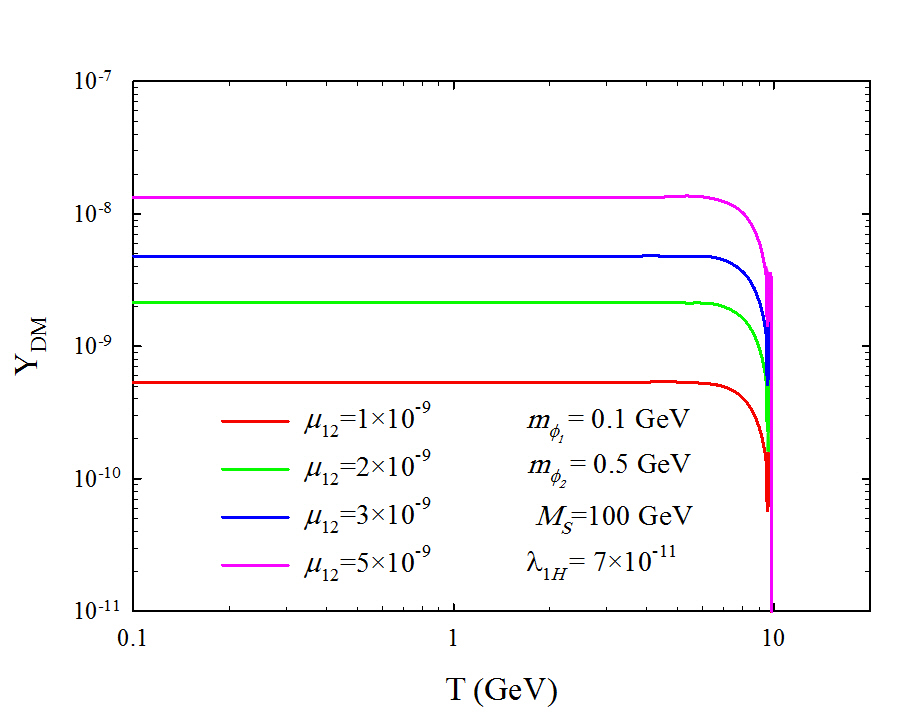}
	\caption{Total DM abundance $Y_{\text{DM}}$ as a function of temperature $T$ for different values of $\mu_{12}$, with $m_{\phi_1}=0.1$~GeV, $m_{\phi_2}=0.5$~GeV, and $M_S=100$~GeV. A log-log scale is adopted for clarity.}
	\label{fig:YvsT_mu12}
\end{figure}

In Fig.~\ref{fig:YvsT_mu12}, we plot the total DM abundance $Y_{\text{DM}} = Y_{\phi_1}^{\text{total}} + Y_{\phi_2}^{\text{total}}$ as a function of the temperature $T$ for four different values of the CP-violating parameter $\mu_{12} = 1\times10^{-9}, 2\times10^{-9}, 3\times10^{-9}, 5\times10^{-9}$, while keeping the other parameters fixed as listed above. The temperature range is chosen to cover the region where the freeze-in production is active.

As expected from the freeze-in mechanism, the final abundance increases with $\mu_{12}$. This is because larger $\mu_{12}$ leads to a larger CP-violating decay width of the mediator $S$, thereby generating a larger asymmetry and a higher DM abundance. For larger values of $\mu_{12}$, the abundance exceeds the observed value, while for smaller values it falls below.

\subsubsection{Dependence on $M_S$}

\begin{figure}[htbp]
	\centering
	\includegraphics[width=0.7\textwidth]{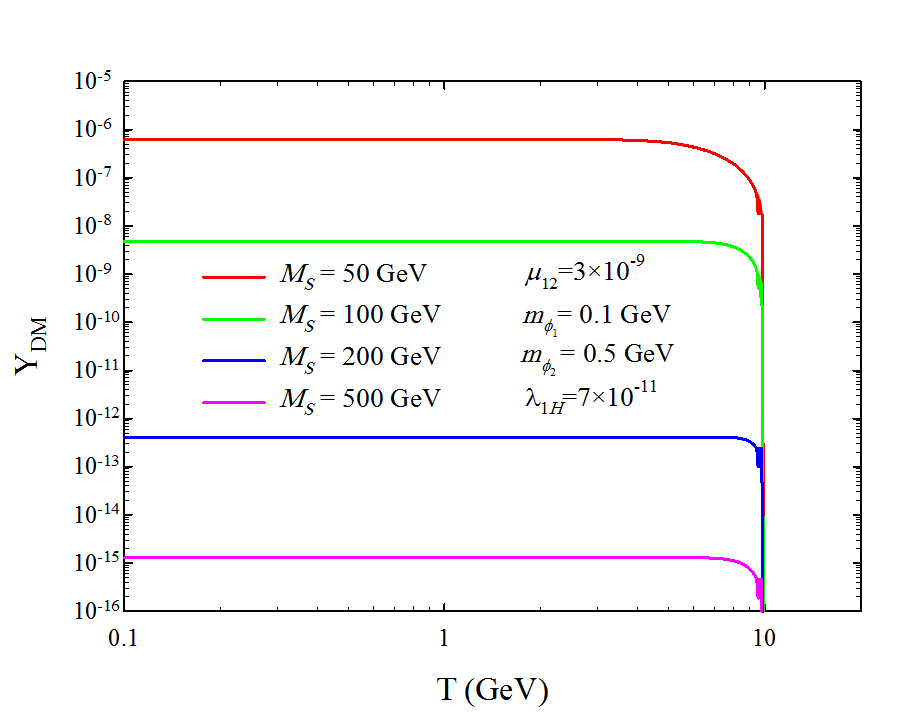}
	\caption{Total DM abundance $Y_{\text{DM}}$ as a function of temperature $T$ for different mediator masses $M_S$, with $m_{\phi_1}=0.1$~GeV, $m_{\phi_2}=0.5$~GeV, and $\mu_{12}=3\times10^{-9}$. A log-log scale is adopted for clarity.}
	\label{fig:YvsT_MS}
\end{figure}

In Fig.~\ref{fig:YvsT_MS}, we show the temperature evolution of $Y_{\text{DM}}$ for four different values of the mediator mass $M_S = 50, 100, 200, 500$~GeV, while fixing $\mu_{12}=3\times10^{-9}$. The effect of the mediator mass is clearly visible: larger $M_S$ leads to a smaller overall abundance. This is because the decay width of the mediator $S$ scales as $\Gamma_{S\to \phi_1\phi_1} \propto 1/M_S$, so heavier mediators produce less DM.

In both figures, the final rise and subsequent plateau of the abundance occur because the production rate is dominated by the decay of the SM Higgs boson, while the contribution from the mediator $S$ is subdominant due to its large mass. Although kinematically accessible for all $T < m_h$, the bath abundance of Higgs bosons becomes exponentially suppressed at $T < m_h/20 \sim 6$ GeV. Hence, the observed flattening of the curves below $T \sim 7$ GeV is a direct signature of the IR-dominated freeze-in regime, where DM production ceases once the temperature drops below the mass of the decaying parent particle.

\subsection{Parameter Space Scan}\label{sec4.2}

We now turn to the relic density $\Omega_{\text{DM}} h^2$ as a function of the free parameters of the model. As before, we fix $m_{\phi_2}=0.5$~GeV, and $\lambda_{1H}=7\times10^{-11}$. The Planck observed value $\Omega_{\text{DM}} h^2 = 0.12 \pm 0.001$~\cite{Planck:2020} is indicated by the gray band in all figures. We focus on the dependence of the relic density on the CP-violating parameter $\mu_{12}$, the mediator mass $M_S$, and the first component mass $m_{\phi_1}$, which are the most relevant parameters for the dynamics of the model.

\subsubsection{Dependence on $\mu_{12}$}

\begin{figure}[htbp]
	\centering
	\includegraphics[width=0.7\textwidth]{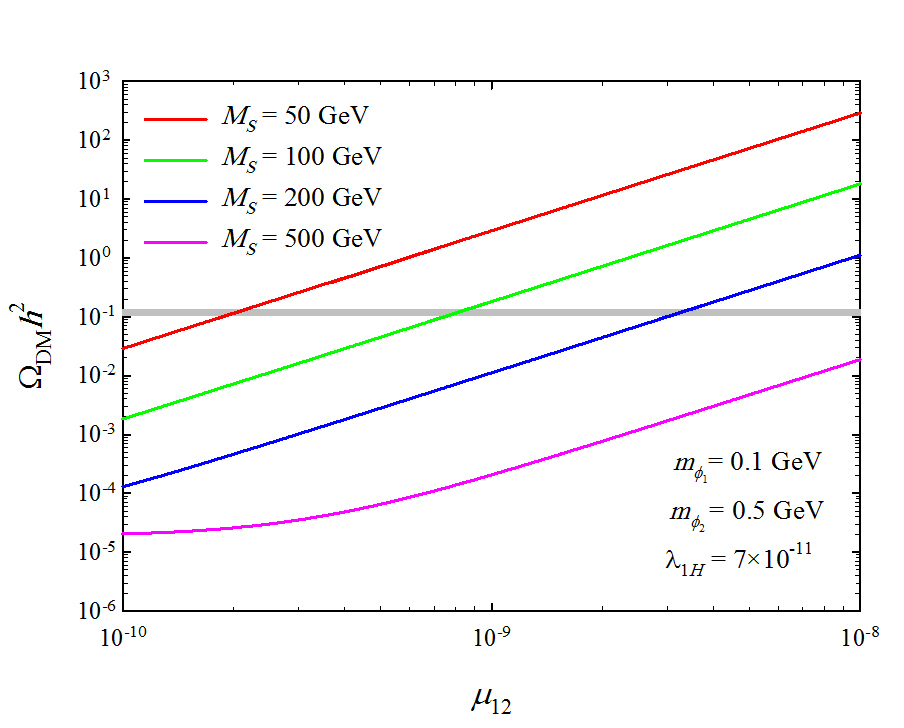}
	\caption{Relic density $\Omega_{\text{DM}} h^2$ as a function of $\mu_{12}$ for different mediator masses $M_S$, with $m_{\phi_1}=0.1$~GeV and $m_{\phi_2}=0.5$~GeV. The gray band indicates the Planck observed range. A log-log scale is adopted to better visualize the parameter dependence.}
	\label{fig:Omega_vs_mu12}
\end{figure}

In Fig.~\ref{fig:Omega_vs_mu12}, we plot the relic density as a function of $\mu_{12}$ for four different values of the mediator mass $M_S = 50, 100, 200, 500$~GeV. As expected, the relic density increases with $\mu_{12}$, since larger $\mu_{12}$ leads to a stronger CP-violating decay of the mediator $S$ and thus a larger asymmetry. For a fixed $\mu_{12}$, larger $M_S$ results in a smaller relic density, as the mediator decay width scales as $\Gamma_{S\to \phi_1\phi_1} \propto 1/M_S$.

The observed relic density $\Omega_{\text{DM}} h^2 \approx 0.12$ is achieved for $M_S = 100$~GeV and $\mu_{12} \approx 3\times10^{-9}$. For smaller values of $\mu_{12}$, the relic density falls below the Planck band, while for larger values it exceeds it. This demonstrates the interplay between the mediator mass and the CP-violating parameter in reproducing the correct DM abundance.

\subsubsection{Dependence on $m_{\phi_1}$}

\begin{figure}[htbp]
	\centering
	\includegraphics[width=0.7\textwidth]{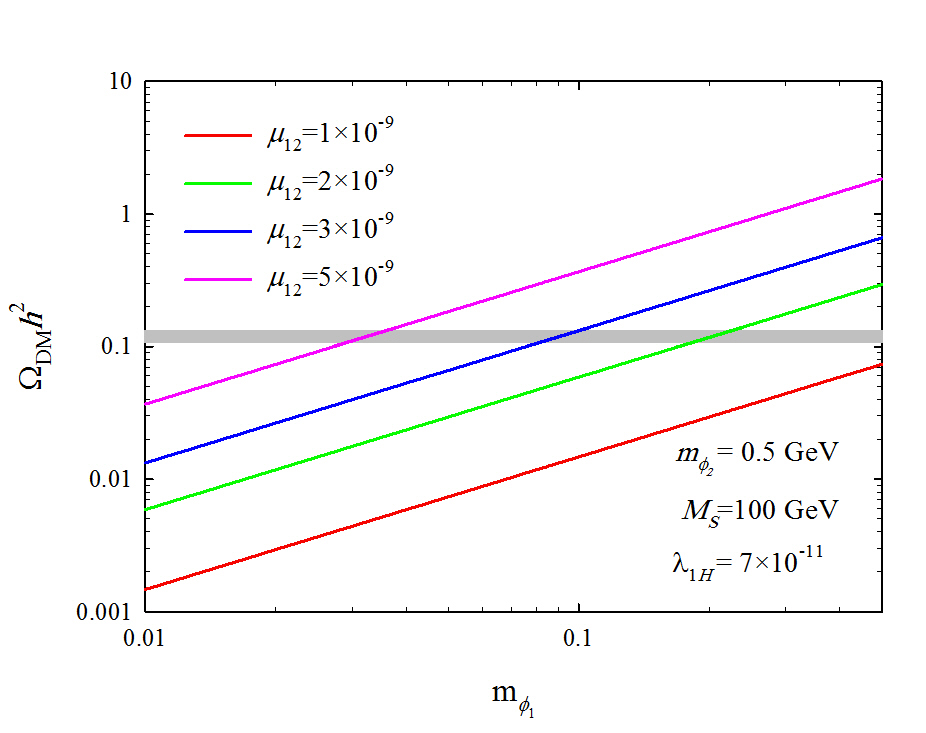}
	\caption{Relic density $\Omega_{\text{DM}} h^2$ as a function of $m_{\phi_1}$ for different values of $\mu_{12}$, with $m_{\phi_2}=0.5$~GeV and $M_S=100$~GeV. The gray band indicates the Planck observed range. Logarithmic scales are used for both axes.}
	\label{fig:Omega_vs_m1}
\end{figure}

In Fig.~\ref{fig:Omega_vs_m1}, we show the relic density as a function of the first component mass $m_{\phi_1}$ for four different values of $\mu_{12} = 1\times10^{-9}, 2\times10^{-9}, 3\times10^{-9}, 5\times10^{-9}$. The results exhibit a clear increasing behavior: larger $m_{\phi_1}$ leads to a larger relic density. This is because the mediator decay width $\Gamma_{S\to \phi_1\phi_1}$ depends on the available phase space, which increases with $m_{\phi_1}$.

For $\mu_{12}=3\times10^{-9}$, the relic density reaches $\Omega_{\text{DM}} h^2 \approx 0.12$ for $m_{\phi_1} \approx 0.1$~GeV. For smaller values of $\mu_{12}$, the relic density remains below the Planck band for the entire range of $m_{\phi_1}$ considered, while for larger $\mu_{12}$ it exceeds the observed value. This confirms the interplay between $\mu_{12}$ and $m_{\phi_1}$ in reproducing the correct DM abundance.

The results confirm that our two-component asymmetric FIMP model can successfully accommodate the observed DM abundance over a wide range of parameters, with the interplay between $\mu_{12}$, $M_S$, and $m_{\phi_1}$ providing the necessary flexibility. In addition, we note that the second component $\phi_2$ contributes only about $5\%$ to the total DM relic density, with the dominant contribution coming from $\phi_1$.


\section{Phenomenological Constraints}\label{sec:pheno}
\subsection{Self-Interacting Dark Matter} \label{sec5.1}
An important phenomenological probe of DM models with feeble couplings to the SM is the self-interaction among DM particles themselves. Unlike direct detection or collider searches, which rely on interactions with SM particles, self-interactions can be substantial even when the portal couplings are extremely small. This makes self-interactions a particularly promising avenue for testing our FIMP scenario.

The most stringent constraints on DM self-interactions come from observations of merging galaxy clusters. The classic constraint from the Bullet cluster, based on the offset between the DM distribution and the galaxy distribution, gives an upper limit $\sigma/m < 0.47$~cm$^2$/g at 95\% C.L.~\cite{Harvey:2015}. More recently, a novel method using double radio relic clusters as merger chronometers has tightened this limit to $\sigma/m < 0.22$~cm$^2$/g at 68\% C.L.~\cite{Jee:2026}. These limits assume velocity-independent cross sections, which is appropriate for our model given the heavy mediator mass.

In our model, the self-interaction of each scalar component arises from the quartic couplings $\lambda_1$ and $\lambda_2$ in the potential. In the non-relativistic limit and for a heavy mediator ($M_S \gg m_{\phi_i}$), the contact interaction dominates, and the self-interaction cross section per unit mass is given by
\begin{equation}
	\frac{\sigma}{m_{\phi_i}} = \frac{\lambda_i^2}{32\pi m_{\phi_i}^3} \times (2.34\times10^{-4})~\text{cm}^2/\text{g},
\end{equation}
where the numerical factor converts from natural units to $\text{cm}^2/\text{g}$.

\begin{figure}[htbp]
	\centering
	\includegraphics[width=0.7\textwidth]{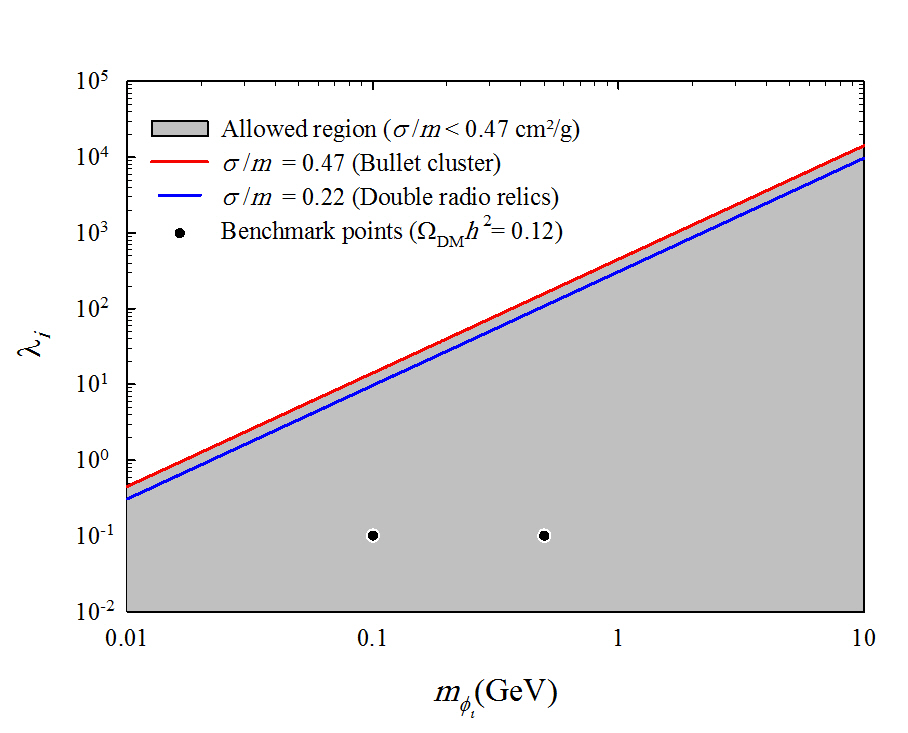}
	\caption{Self-interaction parameter space for both components ($i=1,2$) in the $(m_{\phi_i}, \lambda_i)$ plane. The gray shaded region indicates the allowed parameter space with $\sigma/m < 0.47$~cm$^2$/g. The red and blue lines represent the upper limits $\sigma/m = 0.47$~cm$^2$/g (Bullet cluster) and $\sigma/m = 0.22$~cm$^2$/g (double radio relic clusters), respectively. The black markers indicate the benchmark points of our model, which also satisfy the observed relic density $\Omega_{\text{DM}} h^2 \approx 0.12$.}
	\label{fig:self_interaction}
\end{figure}

To determine the allowed parameter space, we generate a grid of $(m_{\phi_i}, \lambda_i)$ points satisfying $\sigma/m < 0.47$~cm$^2$/g, with the results shown in Fig.~\ref{fig:self_interaction}. Since the self-interaction cross section scales as $\sigma/m \propto \lambda_i^2 / m_{\phi_i}^3$, the allowed parameter space is identical for both components when expressed in terms of $(m_{\phi_i}, \lambda_i)$. Hence, Fig.~\ref{fig:self_interaction} applies universally to $\phi_1$ and $\phi_2$. Therein, the gray shaded region indicates the allowed parameter space ($\sigma/m < 0.47$~cm$^2$/g), while the red and blue lines mark the Bullet cluster limit ($\sigma/m = 0.47$~cm$^2$/g) and the more stringent double radio relic limit ($\sigma/m = 0.22$~cm$^2$/g), respectively.

For our benchmark points, even taking the self-coupling as large as $\lambda_1 = 1$ (which is still within the perturbative regime), the self-interaction cross section for $\phi_1$ is $\sigma/m \sim 2.3\times10^{-3}$~cm$^2$/g, well below both limits. For $\phi_2$, with $\lambda_2 = 1$, we obtain $\sigma/m \sim 1.9\times10^{-5}$~cm$^2$/g. Consequently, these values are several orders of magnitude below the classical and updated bounds. Thus, our model comfortably satisfies all current self-interaction constraints.

\subsection{Other Constraints} \label{sec5.2}
\subsubsection{Invisible Higgs Decay}
The Higgs boson can decay invisibly into pairs of DM particles through the portal couplings $\lambda_{1H}$ and $\lambda_{2H}$. The partial decay width for $h \to \phi_i \phi_i$ is given by
\begin{equation}
	\Gamma_{h \to \phi_i \phi_i} = \frac{\lambda_{iH}^2 v^2}{32\pi m_h} \sqrt{1 - \frac{4m_{\phi_i}^2}{m_h^2}}.
\end{equation}
For our benchmark parameters, with $\lambda_{1H}=7\times10^{-11}$, $\lambda_{2H}=7\times10^{-12}$, and $m_{\phi_i} \ll m_h$, the decay widths are $\Gamma_{h \to \phi_1\phi_1} \approx 2.36\times10^{-20}$~GeV and $\Gamma_{h \to \phi_2\phi_2} \approx 2.36\times10^{-22}$~GeV. The total decay width of the Higgs boson in the SM is $\Gamma_h^{\text{SM}} \approx 4$~MeV, so the invisible branching ratio is
\begin{equation}
	\text{Br}(h \to \text{invisible}) = \frac{\Gamma_{h \to \phi_1\phi_1} + \Gamma_{h \to \phi_2\phi_2}}{\Gamma_h^{\text{SM}} + \Gamma_{h \to \phi_1\phi_1} + \Gamma_{h \to \phi_2\phi_2}} \approx 6\times10^{-18}.
\end{equation}
This is about 16 orders of magnitude below the current experimental upper limits from the LHC, which constrain $\text{Br}(h \to \text{invisible}) < 0.18$ at 95\% C.L.~\cite{ATLAS:2023, CMS:2024}. Therefore, the invisible Higgs decay constraint is automatically satisfied in our model and does not restrict the parameter space.

\subsubsection{Direct Detection}
Direct detection experiments search for elastic scattering of DM particles off nuclei. In our model, the spin-independent scattering cross section is proportional to $\lambda_{iH}^2$ and is further suppressed by the small portal couplings. For the freeze-in regime with $\lambda_{iH} \sim 10^{-11}$--$10^{-12}$, the predicted cross section lies far below the current sensitivity of experiments such as LZ and XENONnT~\cite{LZ:2023, XENON:2023}. Consequently, our model remains consistent with all direct detection constraints.

\section{Conclusion}\label{sec:conclud} 
In this work, we have presented a two-component asymmetric FIMP DM model consisting of two complex scalar fields. These fields are stabilized by a $\mathbb{Z}_2 \times \mathbb{Z}_2'$ symmetry and their abundance is generated via the freeze-in mechanism through the Higgs portal. A heavy scalar mediator is introduced to the dark sector, whose CP-violating decay generates an asymmetry in the first component. Subsequently, this asymmetry is passed to the other component via the quartic coupling between the two scalar fields.

By solving the Boltzmann equations numerically, we have computed the DM relic density and performed a detailed scan over the parameter space. The observed relic density $\Omega_{\text{DM}} h^2 = 0.12 \pm 0.001$~\cite{Planck:2020} is successfully reproduced for the benchmark parameters $m_{\phi_1}=0.1$~GeV, $m_{\phi_2}=0.5$~GeV, $M_S=100$~GeV, and $\mu_{12}=3\times10^{-9}$. In this sense, the dominant contribution to the total relic density comes from $\phi_1$, while $\phi_2$ accounts for only about $5\%$.

We have also examined the phenomenological constraints on the model. The DM self-interaction cross section is found to lie orders of magnitude below the Bullet cluster upper limit of $\sigma/m < 0.47$~cm$^2$/g, and also well below the more stringent double radio relic limit $\sigma/m < 0.22$~cm$^2$/g. The invisible Higgs decay branching ratio is several orders of magnitude below the LHC upper bound. Direct detection prospects are likewise negligible due to the small Higgs portal couplings.

In summary, the two-component complex scalar asymmetric FIMP DM model provides a viable and testable framework that successfully reproduces the observed DM abundance while evading all current experimental constraints. This work establishes a novel connection between two-component DM, the freeze-in mechanism, and DM asymmetry, opening up new avenues for exploring the dark sector.





\end{document}